# Alcoholic beverages induce superconductivity in $FeTe_{1-x}S_x$

K Deguchi[1,2,3], Y Mizuguchi[1,3], Y Kawasaki[1,2,3], T Ozaki[1,3], S Tsuda[1,3], T Yamaguchi[1,3] and Y Takano[1,2,3]

[1] National Institute for Materials Science, 1-2-1, Sengen, Tsukuba, 305-0047, Japan

[2] Graduate School of Pure and Applied Sciences, University of Tsukuba, 1-1-1 Tennodai, Tsukuba, 305-8571, Japan

[3] Japan Science and Technology Agency - Transformative Research-project on Iron-Pnictides (JST-TRIP), 1-2-1, Sengen, Tsukuba, 305-0047, Japan

E-mail: DEGUCHI.Keita@nims.go.jp

Abstract.

We found that hot alcoholic beverages were effective in inducing superconductivity in $FeTe_{0.8}S_{0.2}$. Heating $FeTe_{0.8}S_{0.2}$ compound in various alcoholic beverages enhances the superconducting properties compared to pure water-ethanol mixture as a control. Heating with red wine for 24 hours leads to the largest shielding volume fraction of 62.4 % and the highest zero resistivity temperature of 7.8 K. Some components present in alcoholic beverages, other than water and ethanol, have the ability to induce superconductivity in $FeTe_{0.8}S_{0.2}$ compound.

## 1. Introduction

Since the discovery of Fe-based superconductors, a great deal of study on search for new superconductivity in related compounds has been actively performed [1-6]. The parent phase of Fe-based superconductors basically undergoes an antiferromagnetic transition. To achieve superconductivity, suppression of this antiferromagnetic ordering is needed. Elemental substitutions suppress the antiferromagnetic ordering and produce superconductivity. For example, $BaFe_2As_2$, which is one of the parent phases, becomes superconducting with Ba-, Fe- and As-site substitution: $Ba_{1-x}K_xFe_2As_2$, $Ba(Fe_{1-x}Co_x)_2As_2$ and $BaFe_2(As_{1-x}P_x)_2$ [3,7,8]. Furthermore, superconductivity in $SrFe_2As_2$ is induced by being exposed to water [9]. Since various substitutions can induce superconductivity, search for dopants to induce or enhance superconductivity of Fe-based compounds is an attractive area of study.

FeTe undergoes antiferromagnetic transition around 70 K and does not show superconductivity. Elemental substitution for the Te site can suppress the magnetism. For example, S substitution suppresses the magnetic order, and S-substituted FeTe synthesized using a melting method shows superconductivity [10]. However, the synthesis of superconducting $FeTe_{1-x}S_x$ is difficult owing to the solubility limit caused by a large difference of ionic radius between S and Te. We have reported that $FeTe_{1-x}S_x$ synthesized using a solid-state reaction does not show bulk superconductivity while the antiferromagnetic ordering seems to be suppressed. However, bulk superconductivity is induced in the $FeTe_{1-x}S_x$ sample by the air exposure, water immersion and oxygen annealing [11-13]. Recently we have discovered an amazing method to induce superconductivity. Here we show the inducement of superconductivity in a $FeTe_{0.8}S_{0.2}$ compound by immersing the sample in alcoholic beverages.

## 2. Experimental details

Polycrystalline samples of $FeTe_{0.8}S_{0.2}$ used in this study were prepared using the solid-state reaction method. Powders of Fe, Te and TeS were sealed into an evacuate quartz tube with a nominal composition of FeTe$_{0.8}$S$_{0.2}$, and heated at 600 °C for 10 hours. After furnace cooling, the products were ground, pelletized, sealed into the evacuated quartz tube and were heated again at 600 °C for 10 hours. The pellet was cut into several pieces. Soon after the cutting of the pellet, we immediately carried out the measurement using one of the pieces to obtain the as-grown data. Other pieces (~ 0.15 g) obtained from same pellet were put into a glass bottle (20 ml) filled with alcoholic beverage, beer (Asahi Super Dry, Asahi Breweries, Ltd.), red wine (Bon Marche, Mercian Corporation), white wine (Bon Marche, Mercian Corporation), Japanese sake (Hitorimusume, Yamanaka shuzo Co., Ltd.), shochu (The Season of Fruit Liqueur, TAKARA Shuzo Co., Ltd.) or whisky (The Yamazaki Single Malt Whisky, Suntory Holdings Limited). We also performed a control experiment using a set of samples immersed in pure water, a mixed solution of water and ethanol, and anhydrous ethanol. Although the water-ethanol and alcoholic beverage sets of samples were cut from separate pellets, reproducible results of the control samples, as described below, indicated a small pellet-to-pellet variation. The samples in various liquids were heated at 70 °C for 24 hours. After the heating, samples were taken out from the bottle and the superconducting properties were investigated. The temperature dependence of magnetization was measured using a SQUID magnetometer down to 2 K under a magnetic field of 10 Oe. The shielding volume fraction was estimated from the lowest-temperature value of magnetic susceptibility after zero-field cooling. The electrical resistivity measurements were performed by the standard DC four-terminal method down to 2 K with a current of 1 mA. The typical values of the distance between voltage terminals, width and thickness of the samples were approximately 1, 2 and 2 mm respectively. Powder x-ray diffraction patterns were collected using the $2\theta/\theta$ method with the Cu $K\alpha$ radiation. We confirmed that there was almost no difference in the x-ray pattern between the as-grown sample and the heated samples within the sensitivity of Lab-level x-ray powder diffraction.

**3. Results and discussion**

The temperature dependence of normalized susceptibility for the as-grown $FeTe_{0.8}S_{0.2}$ sample and the samples heated at 70 °C for 24 hours in various water-ethanol mixtures with ethanol concentrations of 0, 20, 40, 60, 80, and 100 % is shown in Fig. 1(a). Firstly, we confirmed that the as-grown sample does not show superconductivity. The samples heated in water-ethanol mixtures at all concentrations showed superconductivity. The estimated shielding volume fractions were 12.4, 8.6, 10.0, 11.1, 9.1, and 5.4 % respectively, and the average value was 9.4 %. We then measured the magnetic susceptibility for the samples heated in beer (ethanol concentration = 5 %), red wine (11 %), white wine (11 %), Japanese sake (15 %), shochu (35 %), and whisky (40 %). The samples heated in these alcoholic beverages also exhibited superconductivity as shown in Fig. 1(b). Surprisingly, the superconducting diamagnetic signals of all the samples heated in alcoholic beverages were clearly larger than that of the samples heated in the water-ethanol mixtures, indicating that the alcoholic beverages are much more effective for evolution of superconductivity in $FeTe_{0.8}S_{0.2}$ than the pure water-ethanol mixture. We estimated the shielding volume fraction of the samples heated in the red wine, white wine, beer, Japanese sake, whisky, and shochu to be 62.4, 46.8, 37.8, 35.8, 34.4, and 23.1 %, respectively; we found that the shielding volume fraction of the red wine sample was the largest and the shochu sample was the smallest. To investigate the reproducibility of these results, we repeated the magnetic susceptibility measurement with the same condition. The shielding volume fractions for the different pellets are almost the same.

The obtained shielding volume fractions are summarized in Fig. 2 as a function of ethanol concentration. The value of the shielding volume fraction for each alcoholic beverage is shown as the mean of seven samples with the standard error, and that for each water-ethanol mixtures is shown as the mean of five samples with the standard error. The shielding volume fractions of the samples heated in water-ethanol mixtures are between 6 and 9 %. In contrast, the shielding volume fraction of the samples heated in alcoholic beverages is 21 - 63 %, significantly larger than that with the water-ethanol mixtures. The value of the sample heated in red wine was more than 6 times larger than the average value of the water-ethanol mixtures. The smallest value among the alcoholic beverages was obtained with shochu, but was still twice larger than the average value of the water-ethanol mixtures. The red wine, white wine, and Japanese sake contain approximately the same ethanol concentration, but they lead to a large difference in the shielding volume fraction. These results suggest that some components of the alcoholic beverages other than hydrous ethanol contribute to the evolution of superconductivity.

Figure 3 is the temperature dependence of normalized resistivity for the as-grown $FeTe_{0.8}S_{0.2}$ sample and the samples heated in beer, red wine, white wine, Japanese sake, shochu, and whisky. The resistivity is normalized at 12 K for comparison. The as-grown sample did not show zero resistivity down to 2 K. Although the onset temperature of the superconducting transition for all the heated samples exhibits almost the same value of 9.9 K, the zero resistivity temperature ($T_c^{zero}$) of the samples slightly depends on the variety of alcoholic beverages. The samples heated in red wine and in white wine show a sharp superconducting transition with $T_c^{zero}$ = 7.8 K. A similar transition is observed for the samples heated in beer, Japanese sake and whisky, showing $T_c^{zero}$ around 7.5 K. The sample heated in shochu exhibits $T_c^{zero}$ around 7.1 K. The samples with a larger shielding volume fraction had a tendency to show a higher $T_c^{zero}$. We found that the sample heated in red wine showed the largest value in both shielding volume fraction and $T_c^{zero}$. And the smallest value among the alcoholic beverages was obtained with shochu, which is highly distilled alcohol, but still clearly larger than that with pure water-ethanol mixtures.

What is the origin of superconductivity induced by the heat treatment in alcoholic beverages? One candidate is the intercalation of ions into the interlayer. There are some reports on superconductivity induced by elementary intercalation [14,15]. If a carrier is generated by the intercalation, superconductivity would be induced. The other candidate to explain the evolution of superconductivity is oxygen in the liquid. In fact, oxygen annealing at 200 °C for the as-grown $FeTe_{0.8}S_{0.2}$ induces bulk superconductivity [13]. However, at lower temperature of 70 °C, hot alcoholic beverages lead to better superconducting properties compared to oxygen annealing at 70 °C. We assume that the alcoholic beverages would play an important role in supplying oxygen into the sample as a catalyst. To elucidate the origin, a detailed analysis of both the structure and composition should be performed.

In conclusion, we found that hot commercial alcoholic beverages were effective in inducing superconductivity in $FeTe_{0.8}S_{0.2}$ compared to pure water, ethanol, and water-ethanol mixtures. The largest shielding volume fraction and the highest $T_c^{zero}$ were achieved by heating the $FeTe_{0.8}S_{0.2}$ sample in red wine. A detailed investigation to clarify the key factor in inducing superconductivity by hot alcoholic beverages is anticipated.

**Acknowledgement**

This work was partly supported by a Grant-in-Aid for Scientific Research (KAKENHI).

**Figure captions**

Fig. 1. Temperature dependence of normalized susceptibility for the as-grown FeTe$_{0.8}$S$_{0.2}$ sample and the samples heated in various liquids. (a) FeTe$_{0.8}$S$_{0.2}$ samples were heated at 70 °C for 24 hours in various water-ethanol mixtures with ethanol concentrations of 0, 20, 40, 60, 80, and 100 %. (b) FeTe$_{0.8}$S$_{0.2}$ samples were heated at 70 °C for 24 hours in various alcoholic beverages.

Fig. 2. The shielding volume fraction of FeTe$_{0.8}$S$_{0.2}$ samples heated in various liquids as a function of ethanol concentration.

Fig. 3. Temperature dependence of normalized resistivity below 12 K for the as-grown sample and the samples heated in the various alcoholic beverages.

Fig. 1(a).

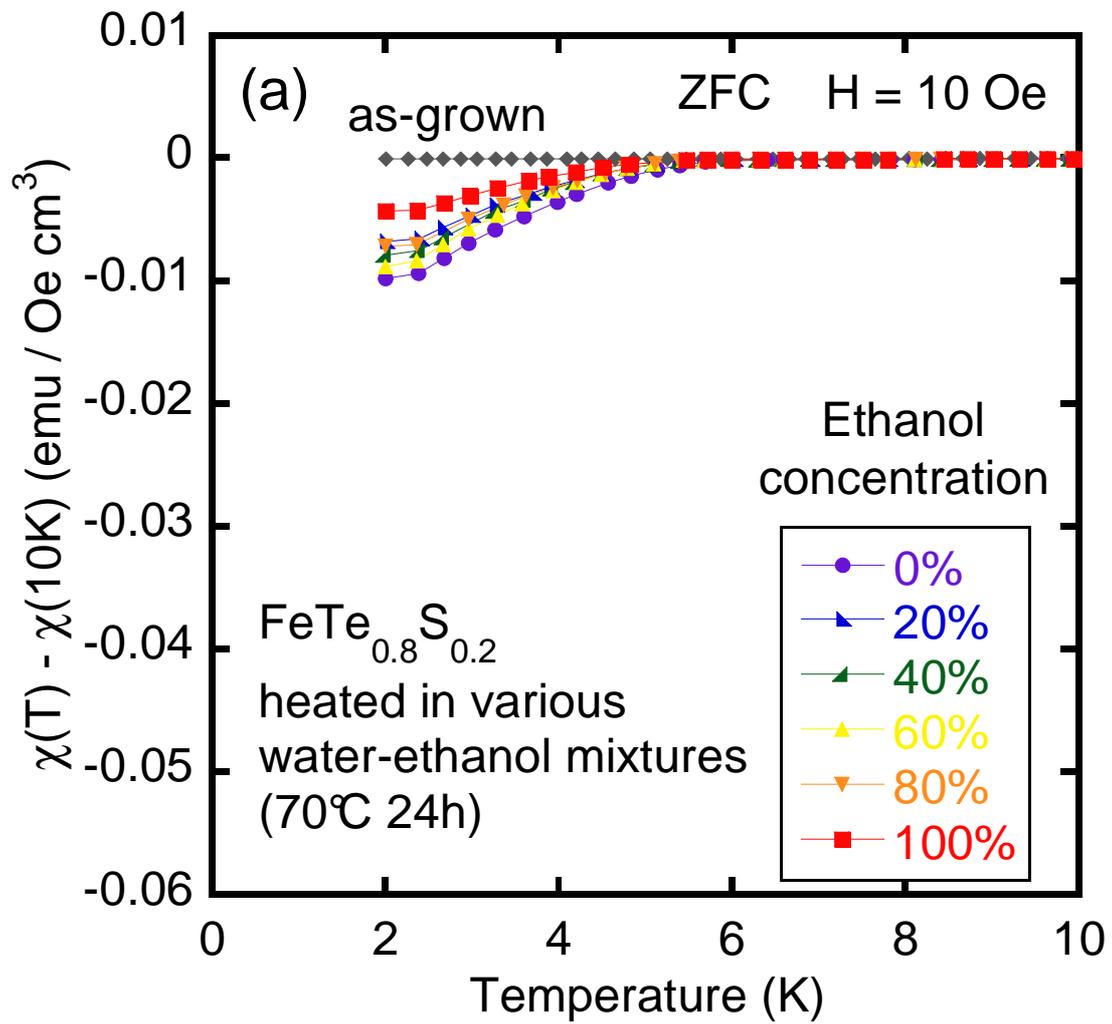

Fig. 1(b).

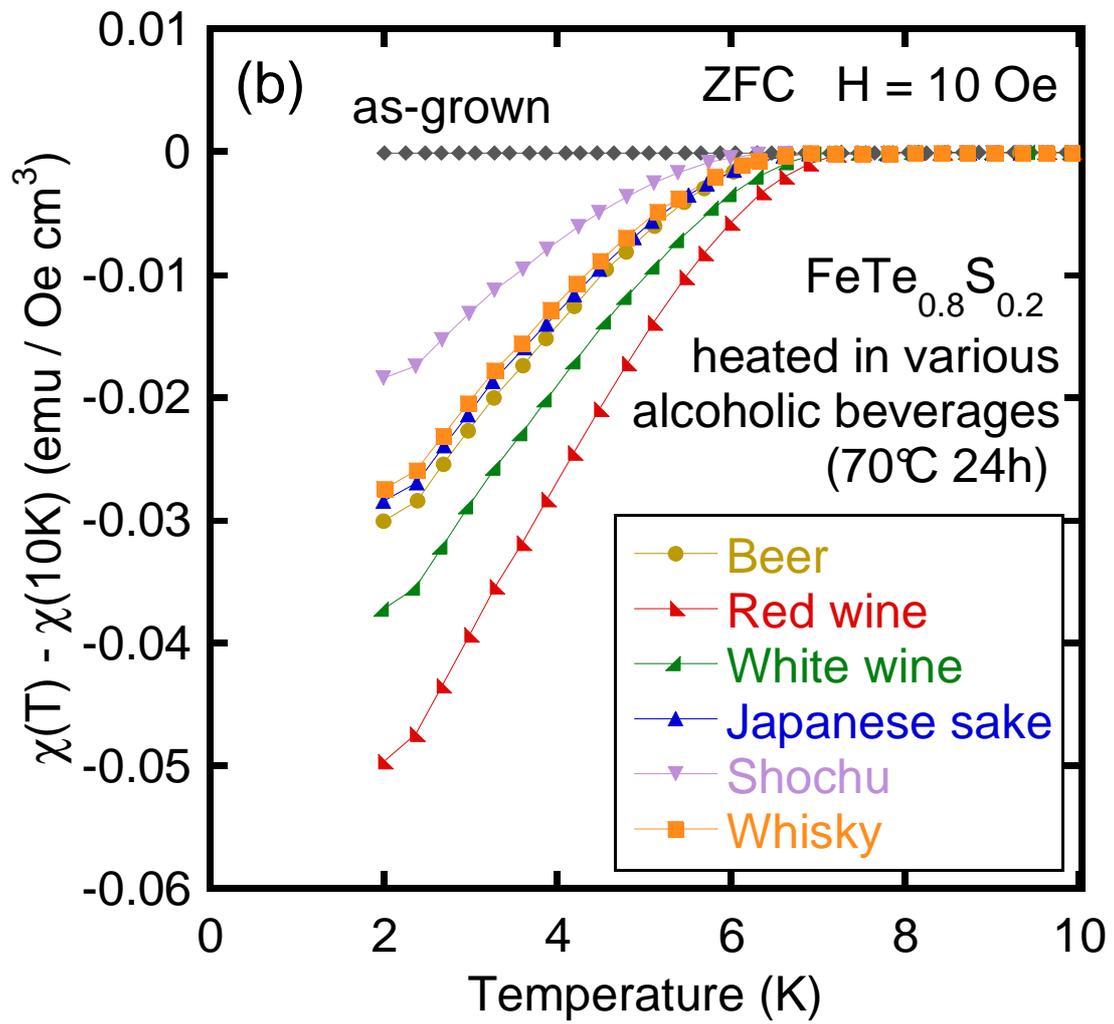

Fig. 2.

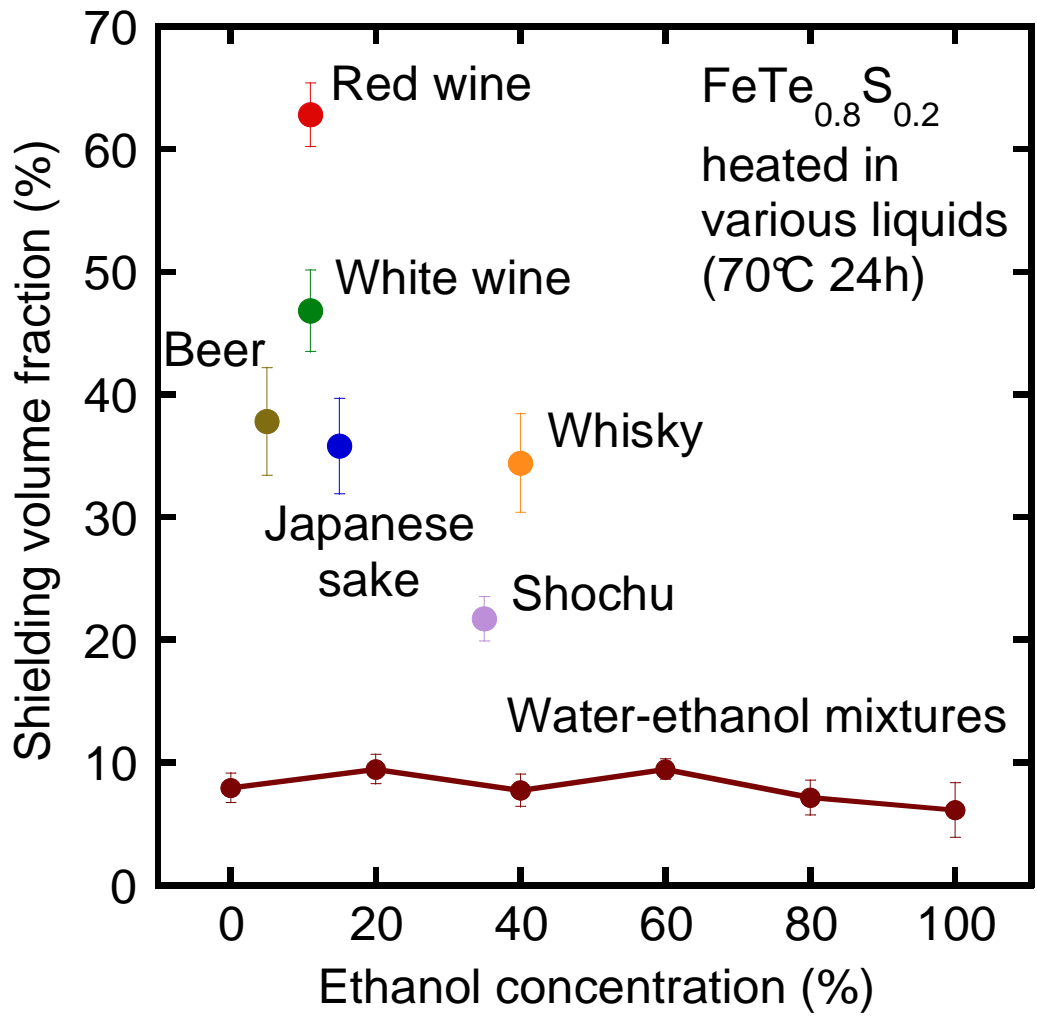

Fig. 3.

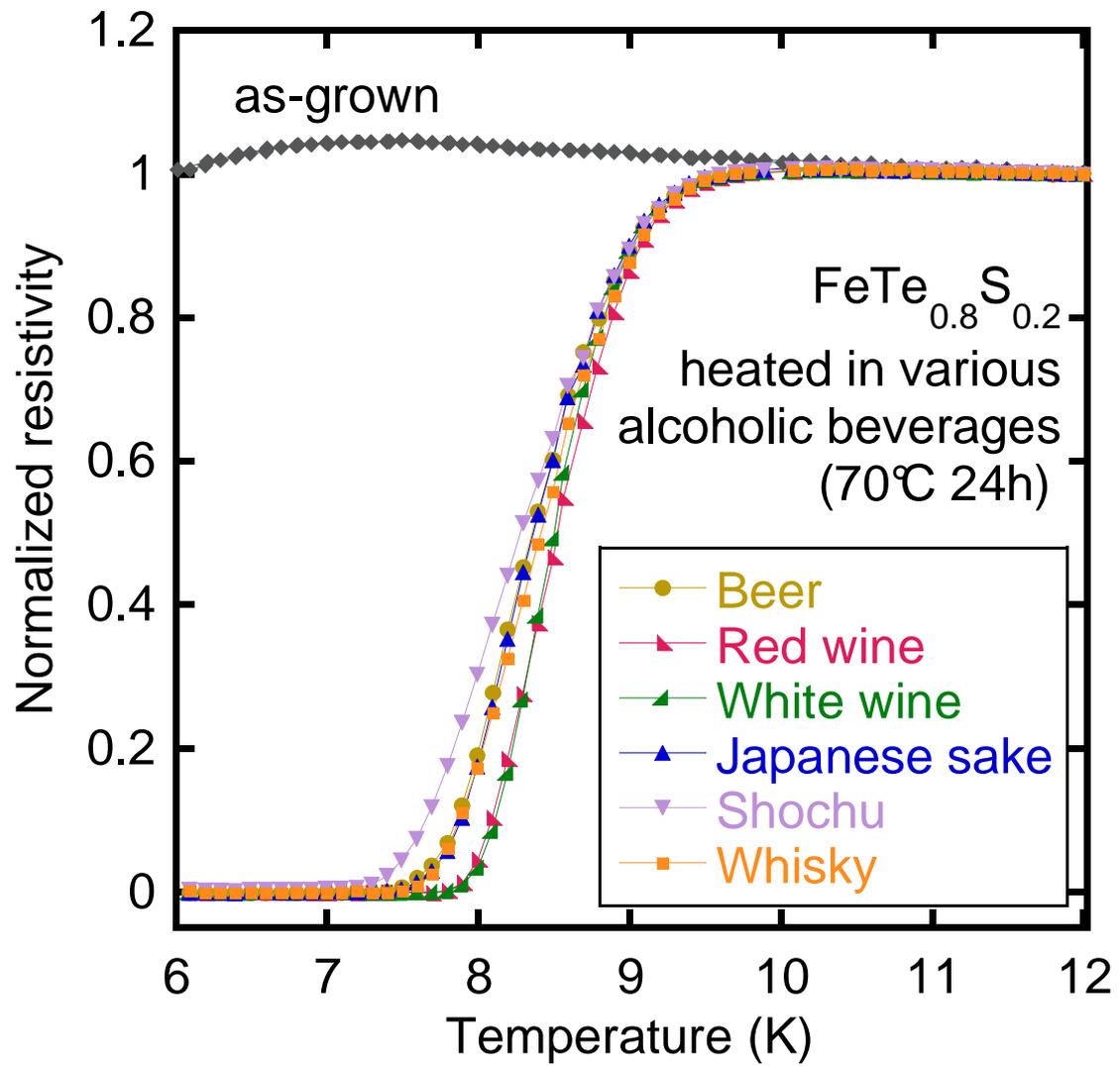